\documentclass[twocolumn, prl, amsmath]{revtex4}
\usepackage{graphicx}
\usepackage{epstopdf}
\newcommand{\mum}[1]{#1\,\mu\text{m}}
\newcommand{\mm}[1]{#1\,\text{mm}}
\newcommand{\nm}[1]{#1\,\text{nm}}
\newcommand{\cm}[1]{#1\,\text{cm}}

\newcommand{\avg}[1]{\left\langle#1\right\rangle}
\newcommand{\vect}[1]{\mathbf{#1}}
\newcommand{\subfig}[1]{(\textbf{#1})}

\newcommand{\vecr}{\vect{r}}
\newcommand{\rk}{\vecr_k}
\newcommand{\rb}{\vecr_b}
\newcommand{\rbeta}{\vecr_\beta}
\newcommand{\ra}{\vecr_a}

\newcommand{\Fourier}[1]{\mathcal{F}\left\{#1\right\}}

\begin{document}
\title{Exploiting disorder for perfect focusing}
\author{I. M. Vellekoop$^{1,2}$, A. Lagendijk$^3$ \& A. P. Mosk$^1$}
\affiliation{
 1. Complex Photonic Systems, Faculty of Science and Technology, and MESA$^+$ Institute for Nanotechnology, University of Twente, P. O. Box 217, 7500 AE Enschede, The Netherlands\\
 2. present address: Physik Institut, University of Zurich, Winterthurerstrasse 190, CH-8057 Zurich, Switzerland\\
 3. FOM Institute for Atomic and Molecular Physics, P. O. Box 41883, 1009 DB Amsterdam, The Netherlands
}

\begin{abstract}
We demonstrate experimentally that disordered scattering can be used to improve,
rather than deteriorate, the focusing resolution of a lens. By using wavefront
shaping to compensate for scattering, light was focused to a spot as small
as one tenth of the diffraction limit of the lens. We show both experimentally
and theoretically that it is the scattering medium, rather than the lens, that
determines the width of the focus. Despite the disordered propagation of the
light, the profile of the focus was always exactly equal to the theoretical best
focus that we derived.
\end{abstract}
\maketitle

Optical microscopy and manipulation methods rely on the ability to focus light to a small volume. However, in inhomogeneous media, such as biological tissue, light is scattered out of the focusing beam. Disordered scattering is thought to fundamentally limit the resolution and penetration depth of optical methods \cite{Ishimaru1978a,Sebbah2001,Hayakawa2009}. Here we demonstrate in an optical experiment that this very scattering can be exploited to improve, rather than deteriorate, the sharpness of the focus. Surprisingly,
the resulting focus is even sharper than in a transparent medium. By using scattering in the medium behind a lens, light was focused to a spot as small as one tenth of the diffraction limit of that lens.
Our results, obtained using spatial wavefront shaping, are valid for all  methods for focusing coherent light through scattering matter, including  phase conjugation\cite{Yaqoob2008} and time-reversal\cite{Derode1995}. We anticipate that disorder-assisted focusing will improve the imaging resolution of microscopy in inhomogeneous media.

\begin{figure}
\centering
  \includegraphics[width=6cm]{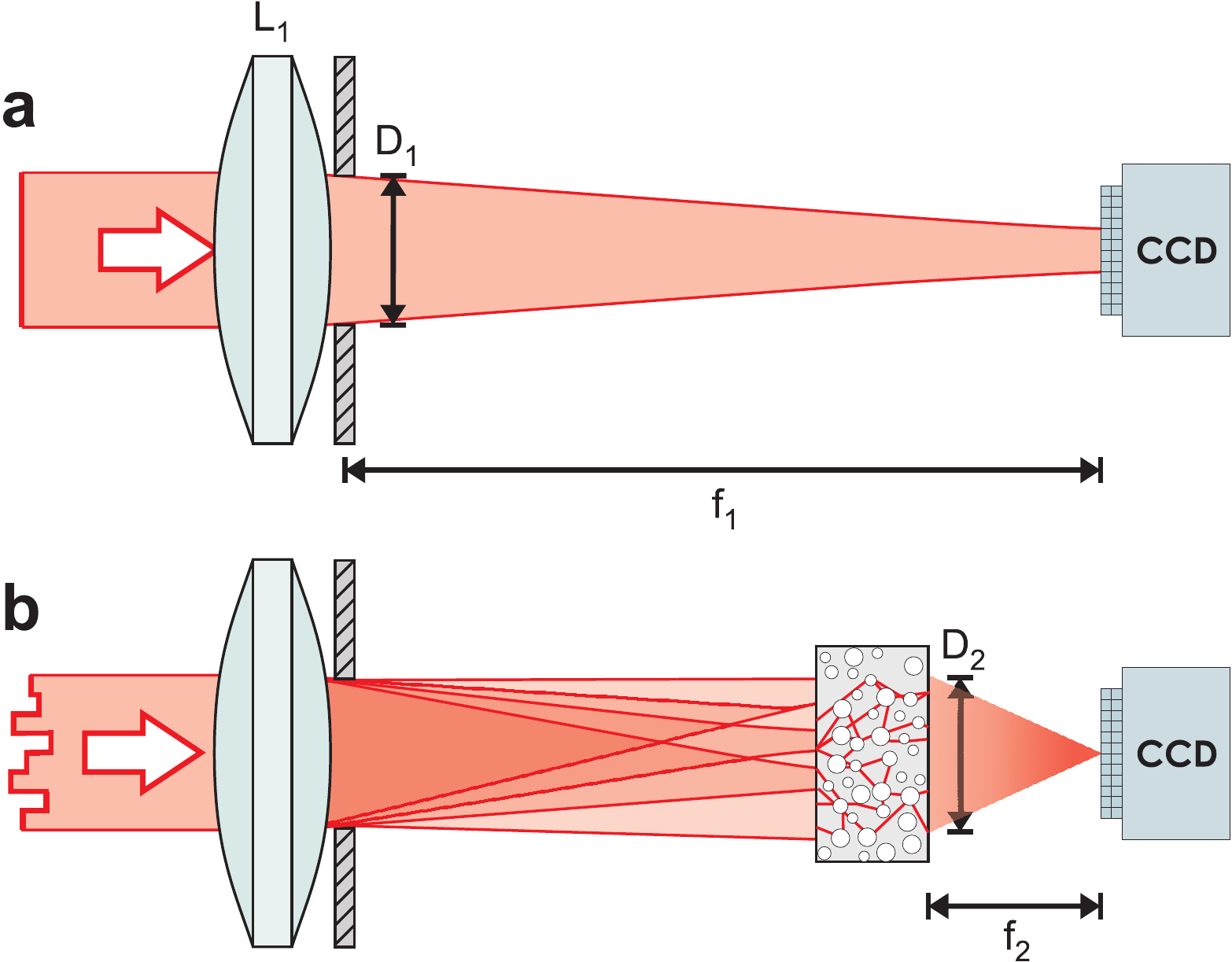}
    \caption{
    Schematic of the experiment. Light coming from a phase modulator is imaged on the centre plane of a lens, $L_1$ (modulator and imaging telescope not shown). The numerical aperture of the lens is controlled by a pinhole. A CCD camera is positioned in the focal plane of the lens. \subfig{a}, `Clean' system without disorder. Light is focused to a spot that is, at best, equal to the diffraction limit of the lens.
    \subfig{b}, System with disorder. A disordered sample randomly changes the direction of the incident light. The scattering object can be moved to change the distance to the camera. When the incident wavefront is shaped to create a focus through the sample, the resulting focus is sharper the best focus the lens can create without disorder.\\[20pt]
    \label{fig:geometry}
    }
\end{figure}

The starting situation of the experiment is shown in Fig.~1a: a lens focuses a beam of light onto a CCD camera. In this `clean' system without disorder, the sharpness of the focus is limited by the numerical aperture and the quality of the lens. We now disturb the light propagation by placing a  non-transparent scattering object in the beam path. Although initially the focus disappears, the focus can be restored by shaping the wavefront of the incident light using a spatial light modulator\cite{Vellekoop2007} (see Fig.~1b).
Here we report and analyze a surprising property of the restored focus: the experimentally obtained focal spot is smaller than the diffraction limit of the clean system. We show both experimentally and theoretically that it is the scattering medium,
rather than the lens or the quality of the reconstruction process, that determines the width of the focus.

\begin{figure}
\centering
  \includegraphics[width=6cm]{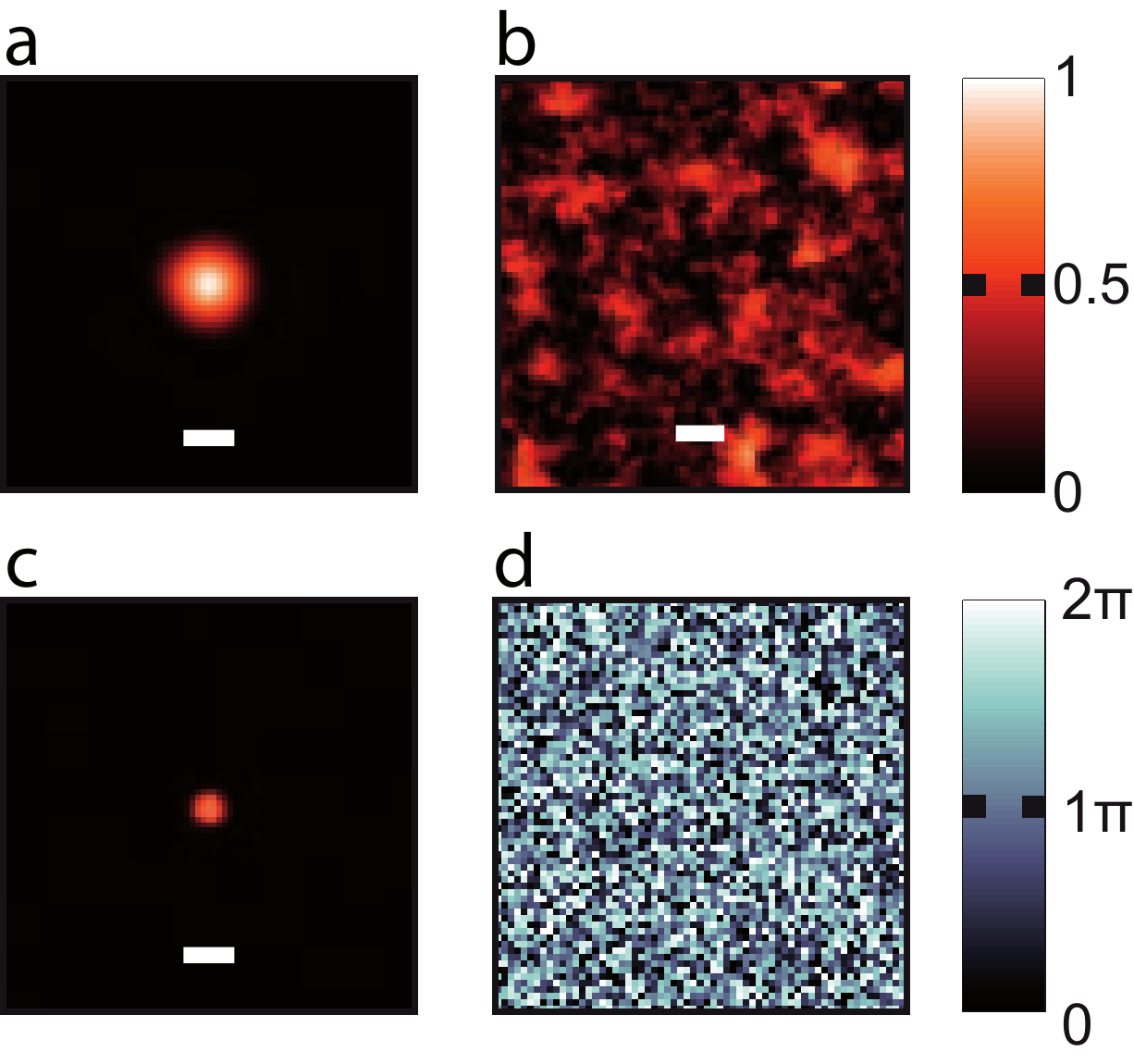}
  \caption{
Measured intensity distribution in the focal plane at $\mm{200\pm3}$ from the glass lens.
  \subfig{a}, Clean system with an unmodified incident wavefront. The focal width is of the order of the diffraction limit ($\mum{62}$, white bar).
  \subfig{b}, System with the $\mum{6}$ layer of airbrush paint present, unmodified incident wavefront. No focus is discernible.
  \subfig{c}, System with the sample present, wave was shaped to achieve constructive interference in the target. A high-contrast, extremely sharp focus is visible.
  \subfig{d}, Pattern on the spatial phase modulator for the situation in Fig.~c.
  The intensity plots are normalized to the brightest point in the image.\\[20pt]
\label{fig:intensityplots}
}
\end{figure}
In Fig.~2a we show the measured intensity distribution in the focal plane of the clean system. Ideally, the lens would focus light to an Airy disk with a full width at half
maximum (FWHM) of $w=1.03\lambda f_1/D_1$. In our experiment,
$\lambda=\nm{632.8}$, $f_1=\mm{200}$, and $D_1=\mm{2.1}$, which
gives a diffraction limited spot size of $\mum{62}$ In reality
\cite{OnlineNote}, the focus has a slightly larger width of $\mum{76\pm3}$.

When the beam path is blocked by a disordered medium (a
$\mum{6}$ layer of opaque white airbrush paint), the image on the camera
changes dramatically. Instead of a focus, we now record a disordered
speckle pattern (see Fig.~2b). Typically, the light is scattered and diffracted about a hundred times before reaching the other side of the sample. Therefore, the transmitted wave has lost all correlation with the incident wavefront\cite{Pappu2002}, and the efficient wavefront
correction systems that have been developed in adaptive optics
cannot be used\cite{Tyson1998}.

We recover the focus using an approach that was designed specifically for strongly
scattering environments. A spatial light modulator shapes the wavefront of the light that impinges on the lens. The surface area of the light modulator was subdivided into $64\times64$ square segments, which are phase-modulated\cite{Putten2008} and controlled by a learning feedback algorithm\cite{OnlineNote}. The algorithm adjusts the relative phases
of the segments so that the transmitted light interferes constructively in a chosen target, thereby creating a focus at that
point. In this case, the target was formed by a single
$\mum{6.45}\times\mum{6.45}$ CCD pixel in the centre of the desired focus. The algorithm finds the optimal wavefront for focusing on this target (see Fig.~2d). The optimal wavefront is completely disordered, which confirms that the sample is strongly scattering.

In Fig.~2c we show the intensity distribution after running the
algorithm. The scattered light now focuses to a tight spot, even though the algorithm did not optimize the shape of the focus explicitly. What is
more, the width of this spot is approximately one tenth of the
diffraction limit of the lens: scattering has greatly improved the focusing
resolution.

To analyze this striking effect, we placed the sample at different distances $f_2$ from the camera. In Fig.~3 we show the width of the focus in these experiments. The widths decrease as the sample is moved closer to the camera, and are always smaller than the width of the diffraction limited spot of the clean system. At distances of $\mm{25}$ or smaller, the width of the focus is smaller than a single camera pixel, almost a factor ten below the diffraction limit of the lens.

We proceed to quantitatively define an effective diffraction limit for a
scattering system. We will show below that the width of the smallest
achievable focus is given by the FWHM of an Airy disc
\begin{equation}
w=1.03\lambda f_2 / D_2,
\label{eq:Effective-Diffraction-Limit}
\end{equation}
with $f_2$ the distance between the scattering layer and the focal plane and
$D_2$ the diameter of the illuminated area on the scattering layer\cite{OnlineNote}.

\begin{figure}
\centering
  \includegraphics[width=6cm]{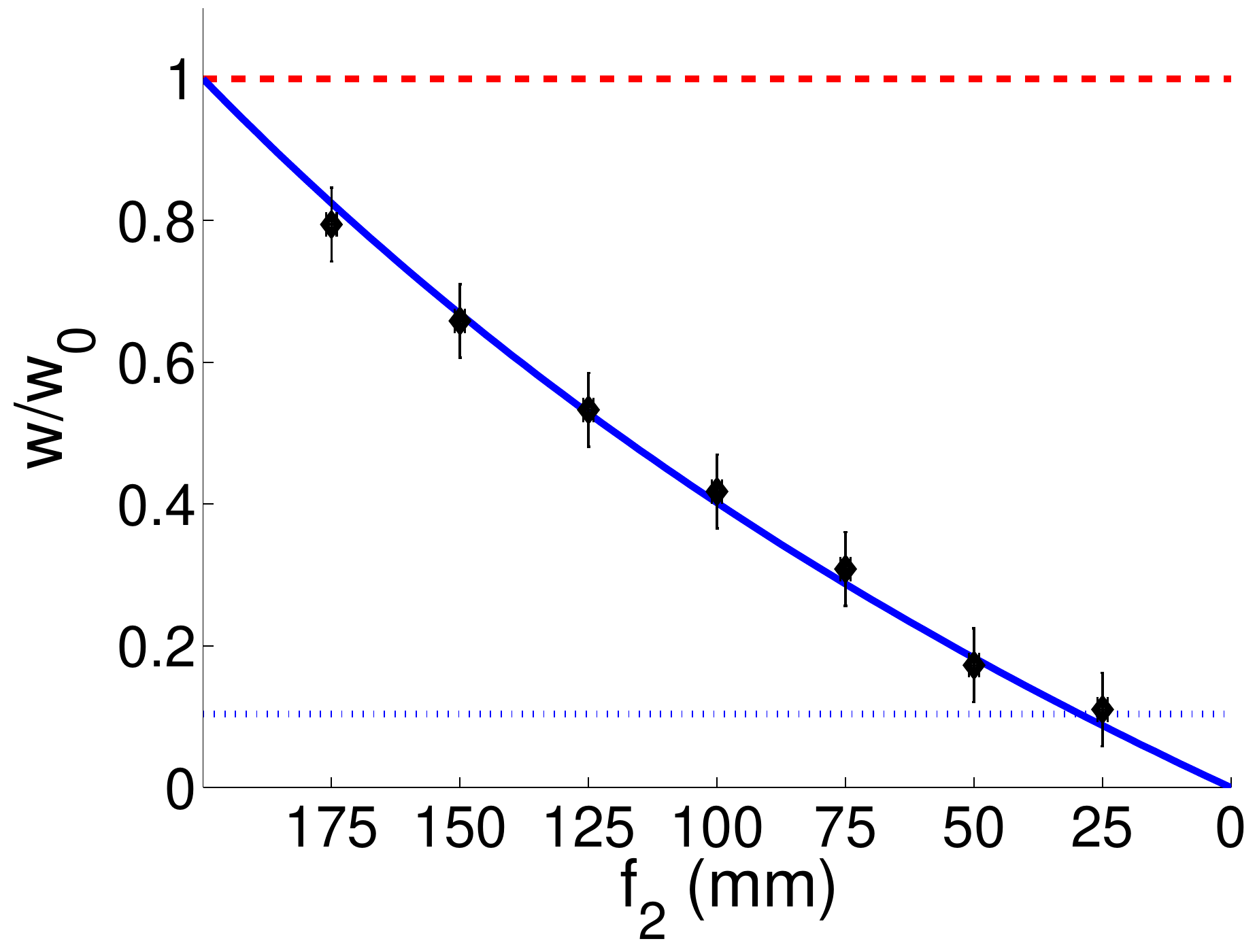}
    \caption{
%
    Focal width $w$, relative to the diffraction limit of the glass lens ($w_0=\mum{62}$, dashed line), as
     a function of the distance $f_2$ between the scattering sample and the focal plane.
     Diamonds, measured values with a sample present; solid curve, theoretical
        effective diffraction limit (no free parameters); dotted line, size of a single pixel of the camera.\\[20pt]
\label{fig:z-scan}
}
\end{figure}

In Fig.~3 we plot our theoretical effective diffraction limit as a
function of the distance between the scattering layer and the focal
plane. The theoretical limit decreases almost linearly as the sample
is moved closer to the chosen focus. With no adjustable parameters, the experimentally observed focal widths  equal the theoretical lower limit that our model predicts.

To understand why the scattered light focuses to exactly the
theoretically smallest possible focus - even though no effort is made to explicitly minimize the width of the focus - we now calculate how a shaped wavefront
propagates through a scattering sample. The propagation of light
from a source at $\ra$ to a target at $\rbeta$ through a sample is
described by the stochastic Green function $G(\rbeta, \ra)$.
Theoretically\cite{Tanter2000,Vellekoop2008c}, the maximum intensity
transmission is achieved when the incident field $E(\ra)$ is
modulated spatially so that $E(\ra)\propto G^*(\rbeta, \ra)$, where $\;^*$ denotes the complex conjugate. Our algorithm constructs this wavefront by maximising the intensity at point $\rbeta$. This is the same wavefront that one would get when one would phase conjugate the light\cite{Yaqoob2008} coming from a (hypothetical) source at $\rbeta$.

If all channels on both sides of the sample could be controlled, the light would form a perfect spherical wave\cite{Tanter2000} that converges to point $\rbeta$.
In a transmission geometry, only a finite number of channels on one side of the sample are controlled. Therefore it is, even in principle, impossible to perfectly focus
all incident light. However, since the sample completely scrambles the incident wavefront, any deviation from the perfect wave is randomly distributed over all outgoing angles. Therefore, the average field at the back of the sample still constitutes a converging spherical wave originating from the illuminated part of the sample\cite{OnlineNote}
\begin{equation}
\avg{E(\rk;\rbeta)} = g^*(\rbeta-\rk) C_0 I_0(\rk)\label{eq:Eoptrk},
\end{equation}
where $E(\rk;\rbeta)$ denotes the field at point $\rk$ on the back of the sample after programming the modulator to optimize the intensity in target $\rbeta$. $g$ is the Green
function for propagating through air, $C_0$ is a constant to normalize the total incident power, and $I_0(\rk)$ is the ensemble averaged intensity distribution at the back of the sample.
Equation~\ref{eq:Eoptrk} shows that experimental limitations do not change the shape of the focus, they merely cause the contrast with the speckle background to decrease.

The term $I_0(\rk)$ in Eq.~\ref{eq:Eoptrk} acts as an effective aperture function for the scattering `lens'. In our experiments $I_0(\rk)$ is approximated well by a top hat disk with
a diameter of $D_2$. In this case, Eq.~\ref{eq:Eoptrk} describes
diffraction limited focusing at point $\rbeta$ by a lens with a
clear aperture of $D_2$. This result explains the experimental observation in Fig~3 that the focal width always equals the theoretical minimum.

The intensity distribution in the focal plane is found from
Eq.~\ref{eq:Eoptrk} by paraxially propagating the field. We derived\cite{OnlineNote}
\begin{equation}
I(\rb;\rbeta) = S_I |\Fourier{I_0(\rk)}|^2 \label{eq:Ioptrb},
\end{equation}
with $\Fourier{}$ the two-dimensional Fourier transform, and $S_I$ a normalization constant that is fixed by the condition that $I(\rbeta;\rbeta)=I_0(\rbeta)(\pi(N-1)/4+1)$, with $N$ the number of
independent modulator segments and $I_0$ the
average intensity of the diffuse background before
optimization\cite{Vellekoop2007,Vellekoop2008c}. The intensity in the focus increases linearly with the number of segments, until $N$ is equal to the total number of degrees of freedom of the optical field.

Equation~\ref{eq:Ioptrb} predicts the shape of the reconstructed focus.
Surprisingly, this equation takes the same form as the van
Cittert-Zernike theorem\cite{Cittert1934,Zernike1938,Goodman2000}. This theorem tells us that the speckle correlation function is proportional to $\left|\Fourier{ I_0(\rk)}\right|^2$.
Our derivations suggest an alternative interpretation of this decades-old theorem: the theorem predicts the exact shape to which a disordered sample will focus an optimized wavefront.

\begin{figure}
\centering
  \includegraphics[width=6cm]{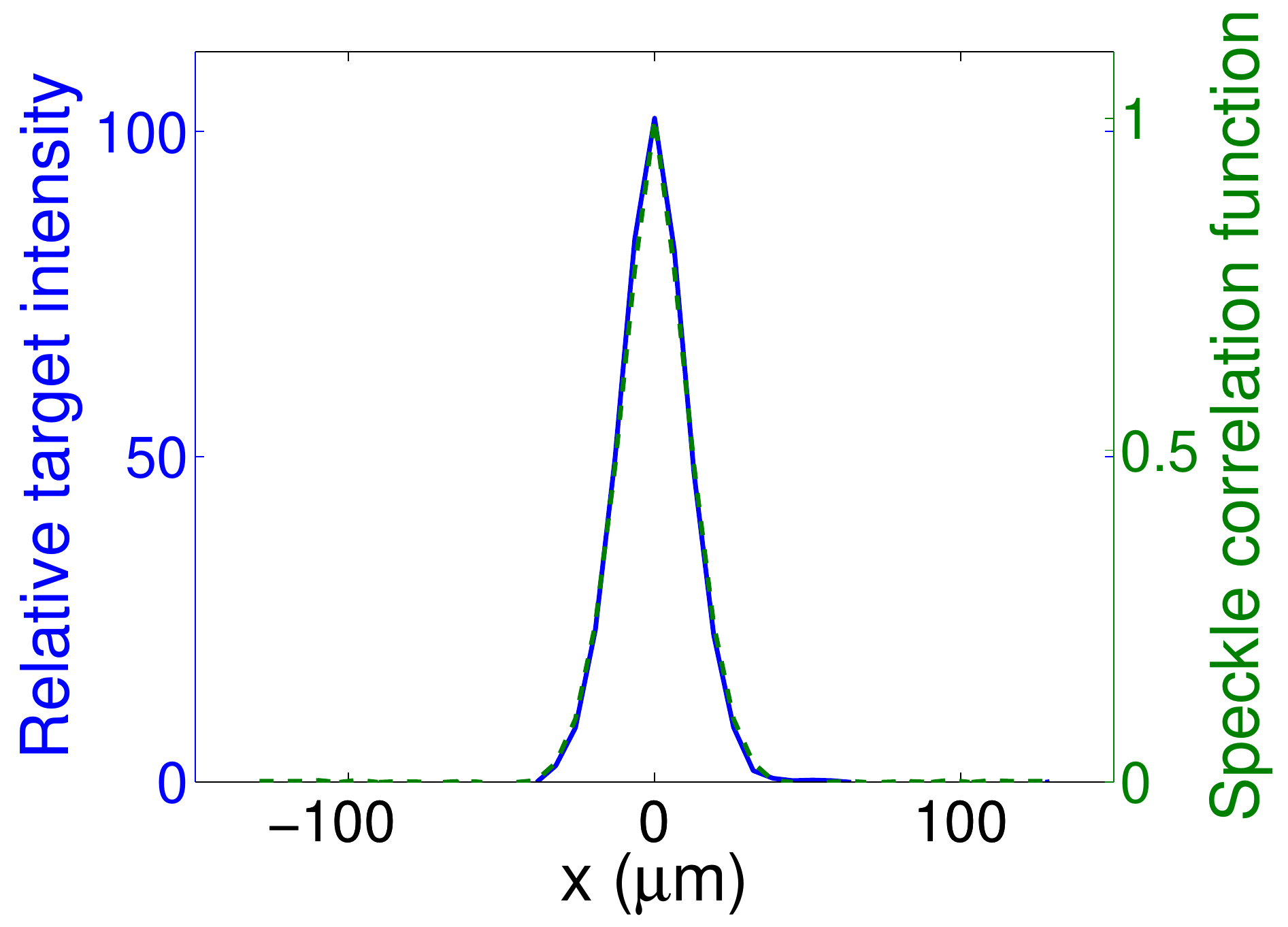}
\caption{
%
  Intensity profile of the focus at $y=0$ (solid curve) and speckle correlation function (dashed curve)
  for an $\mum{6}$ thick layer of airbrush paint creating a focus at $\cm{10.0}$ distance. The speckle correlation function
  was measured with a randomly generated incident wavefront.
  }
\end{figure}
To test this remarkable result, we compare the measured shape of the reconstructed focus to the measured autocorrelate of the background speckle. In Fig.~4 the two normalised curves can be seen to overlap perfectly. We repeated this experiment in different geometries and with different samples and found that the shape of the focus is always in agreement with the shape that the van Cittert-Zernike theorem predicts.

In conclusion, scattering in a medium behind a lens can be used to
\emph{improve} the focusing resolution to beyond the diffraction
limit of that lens.
 We found that, surprisingly, the shape of the
focus is not affected by experimental limitations of the wavefront
modulator: the focus is always exactly as sharp as is theoretically
possible.

Disordered scattering has been applied to improve resolution and bandwidth in imaging and
communication with ultrasound, radio waves and microwaves\cite{Derode1995,Foschini1996,Simon2001}, and significant sub-wavelength effects have been demonstrated \cite{Lerosey2007}.
Our results are the first demonstration that similar resolution improvements can be
obtained in photonics. Calculations \cite{Stockman2002,Bartal2009} indicate that useful optical superresolution will be achieved using disordered plasmonic nanostructures.


\begin{description}
 \item[Acknowledgments] We thank Willem Vos and Elbert van Putten for support and discussions.
 This work is part of the research program of
the ``Stichting voor Fundamenteel Onderzoek der Materie (FOM)",
which is financially supported by the ``Nederlandse Organisatie voor
Wetenschappelijk Onderzoek (NWO)". A.~P.~Mosk is supported by a VIDI
grant from NWO.
\end{description}

\end{document}